\documentclass{ifacconf}

\counterwithin*{section}{part}

\usepackage{graphicx}      
\usepackage{natbib}        

\usepackage{amsmath,amssymb,amsfonts}
\usepackage{verbatim}   
\usepackage{algorithm}
\usepackage{algpseudocode}

\usepackage{mathtools}
\usepackage{siunitx}
\usepackage{tabularx}
\usepackage{pgfplots}
\usepackage{pgfplotstable}
\usepackage{multirow}
\usepgfplotslibrary{fillbetween}
\usepackage{booktabs}
\pgfplotsset{compat = newest}
\usetikzlibrary{arrows,positioning,shapes,intersections,external,patterns,calc,fit,decorations,decorations.markings,spy} 
\usepackage{balance}

\newcommand\tran{\mkern-2mu\raise1.25ex\hbox{$\scriptscriptstyle\top\hspace{0.5mm}$}\mkern-3.5mu}
\newcommand{\R}{\mathbb{R}}

\newcommand{\N}{\mathbb{N}}
\newcommand{\C}{\mathcal{C}}

\newcommand{\D}{\mathcal{D}}
\newcommand{\X}{\mathcal{X}}

\newcommand{\bm}[1]{{\boldsymbol{#1}}}

\DeclareMathOperator{\diag}{diag}
\DeclareMathOperator{\var}{var}

\DeclareMathOperator{\Prob}{P}
\DeclareMathOperator{\proby}{p}
\newcommand{\GP}{\mathcal{GP}}

\newcommand{\x}{\bm x}

\newcommand{\dx}{\dot{\bm x}}

\renewcommand{\u}{\bm{u}}
\newcommand{\y}{\bm{y}}

\usepackage[noabbrev]{cleveref} 
\crefname{rem}{Remark}{Remarks}
\crefname{exam}{Example}{Examples}
\crefname{assum}{Assumption}{Assumptions}
\crefname{prop}{Proposition}{Propositions}
\crefname{propy}{Property}{Properties}
\crefname{cor}{Corollary}{Corollaries}
\crefname{lem}{Lemma}{Lemmas}
\crefname{section}{Section}{Sections}
\crefname{thm}{Theorem}{Theorems}
\crefname{defn}{Definition}{Definitions}
\crefname{figure}{Fig.}{Fig.}
\Crefname{figure}{Figure}{Figures}
\crefname{equation}{}{}
\begin{document}
\begin{frontmatter}

\title{Learning Switching Port-Hamiltonian Systems with Uncertainty Quantification} 


\author[First]{Thomas Beckers} 
\author[Second]{Tom Z. Jiahao} 
\author[Third]{George J. Pappas} 

\address[First]{Department of Computer Science, Vanderbilt University, Nashville, TN 37212, USA, (e-mail: thomas.beckers@vanderbilt.edu)}
\address[Second]{Department of Computer and Information Science, University of Pennsylvania, Philadelphia, PA 19104, USA, (e-mail: zjh@seas.upenn.edu)}
\address[Third]{Department of Electrical and Systems Engineering, University of Pennsylvania, Philadelphia, PA 19104, USA, (e-mail: pappasg@seas.upenn.edu)}

\begin{abstract}
Switching physical systems are ubiquitous in modern control applications, for instance, locomotion behavior of robots and animals, power converters with switches and diodes. The dynamics and switching conditions are often hard to obtain or even inaccessible in case of a-priori unknown environments and nonlinear components. Black-box neural networks can learn to approximately represent switching dynamics, but typically require a large amount of data, neglect the underlying axioms of physics, and lack of uncertainty quantification. We propose a Gaussian process based learning approach enhanced by switching Port-Hamiltonian systems (GP-SPHS) to learn physical plausible system dynamics and identify the switching condition. The Bayesian nature of Gaussian processes uses collected data to form a distribution over all possible switching policies and dynamics that allows for uncertainty quantification. Furthermore, the proposed approach preserves the compositional nature of Port-Hamiltonian systems. A simulation with a hopping robot validates the effectiveness of the proposed approach.

\end{abstract}


\end{frontmatter}

\section{Introduction}
The modeling and identification of switching dynamical systems is a crucial task in a wide range of domains, such as robotics and power systems~\citep{anderson2020model, brogliato1999nonsmooth, wu2018switched}. System identification has a long history in control as many control strategies are derived based on a precise model of the plant. Whereas classical models of physical systems are typically based on first principles, there is a recent interest in data-driven modeling of switching systems to capture more details with reduced engineering effort.

However, this paradigm shift poses new questions regarding the efficiency, interpretability, and physical correctness of these models~\citep{hou2013model}. By physical correctness, we mean that the learned model respects physical principles such as the conservation of energy and passivity. Including these physical principles in a data-driven approach is beneficial in several ways: The models i) are more meaningful as they respect the postulates of physics, ii) come with increased interpretability, and iii) can be more data-efficient, since the satisfaction of physical axioms results in a meaningful inductive bias of the model \citep{karniadakis2021physics}. A large class of switching physical systems can be described by Port-Hamiltonian systems. 
\begin{figure}[t]
\begin{center}
\vspace{0.0cm}
	\input{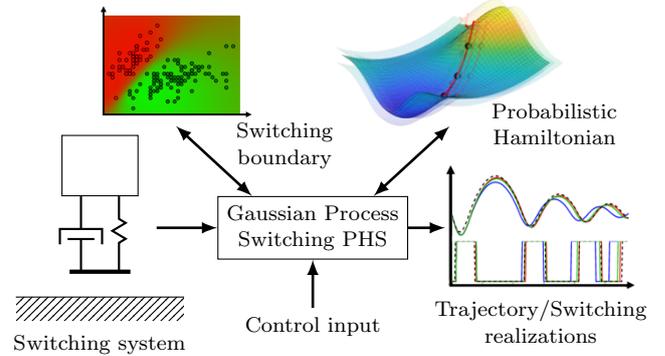}
	\vspace{-0.2cm}\caption{Gaussian Process switching Port-Hamiltonian systems (GP-SPHS) allow to learn a probabilistic Hamiltonian and switching policy based on collected data. All sample trajectories of GP-SPHS are physically correct in terms of the evolution of the energy.}\vspace{-0.1cm}
	\label{fig:gpphs_intro}
\end{center}
\end{figure}
Port-Hamiltonian systems (PHS) leverage network approaches from electrical engineering and constitute a cornerstone of mathematical systems theory. While most of the analysis of physical systems has been performed within the Lagrangian and Hamiltonian framework, the network point of view is attractive for the modeling and simulation of complex physical systems with many connected components. The main idea of switching Port-Hamiltonian systems (SPHS) is that the total energy in the system is described by the Hamiltonian, a smooth scalar function of the system's generalized coordinates which is not affected by any switching~\citep{van2014port}. Instead, the switching is modeled by a rerouting of the energy flow that may be internally induced (state dependent) or externally triggered. It has been shown that SPHS are able to model a range of multi-domain complex physical systems, in particular, mechanical and electrical systems~\citep{van2000l2}. However, the Hamiltonian and the switching policy are often highly nonlinear and, thus, challenging for modeling and identification. The purpose of this paper is to combine the expressiveness of Gaussian processes with the physical plausibility of switching Port-Hamiltonian systems to model physical switching systems with unknown dynamics and switching condition based on data. 

There is a rapidly growing body of literature on data-driven modeling of dynamical systems using deep learning. In particular, switching dynamics arises in stiff systems or systems with frictional contact~\citep{anderson2020model}. Notably, a family of neural networks, neural ordinary differential equations, was proposed to represent the vector field of dynamical systems using neural networks~\citep{conf/nips/ChenRBD18}, but the successes have been limited to low-stiff systems~\citep{jiahao2021knowledgebased}. Efforts to build models that conserve the total energy of a system led to a body of work on Hamiltonian neural networks (HNN)~\citep{greydanus2019hamiltonian}. However, since many physical systems are dissipative, the assumption of HNNs limits its applicability to real-world systems. Results for data-driven learning of PHS with neural networks have been established in~\citep{desai2021port,nageshrao2015port}. However, standard neural network architectures can have difficulty generalizing from sparse data and have no uncertainty quantification of the outputs. In contrast, Bayesian models such as Gaussian processes (GP) can generalize well even for small training datasets and have a built-in notion of uncertainty quantification~\citep{rasmussen2006gaussian}. GPs have been recently used for learning generating functions for Hamiltonian systems~\citep{rath2021symplectic}, dynamics of structured systems~\citep{ridderbusch2021learning,bhouri2021gaussian} and PHS~\citep{9992733}, but they have not been applied to the class of switching PHS.

\textbf{Contribution:}  We propose a Bayesian learning approach using Gaussian processes and SPHS to learn switching physical systems with partially unknown dynamics based on data.
In contrast to existing modeling methods for SPHS, the probabilistic model includes all possible realizations of the SPHS under the GP prior, which guarantees physical plausibility, high flexibility, and allows for uncertainty quantification. We prove that the passivity and interconnection properties of SPHS are preserved in GP-SPHS, which is a beneficial characteristic for the composition and control of physical system.

The remainder of the paper is structured as follows. We start with the introduction of SPHS in~\cref{sec:sphs} and the problem setting in~\cref{sec:ps} followed by proposing GP-SPHS in~\cref{sec:mod}. Finally, a simulation highlights the benefits of GP-SPHS in~\cref{sec:sim}.
\section{Switching Port-Hamiltonian Systems}\label{sec:sphs}
Composing Hamiltonian systems with input/output ports leads to a Port-Hamiltonian system, which is a dynamical system with ports that specify the interactions of its components. The switching of the dynamics can occur due to an internal or external trigger which affects the routing of the energy~\citep{van2009state}. The dynamics of a SPHS is described by\footnote{Vectors~$\bm a$ and vector-valued functions~$\bm f(\cdot)$ are denoted with bold characters. Matrices are described with capital letters. $I_n$ is the $n$-dimensional identity matrix and $0_n$ the zero matrix. The expression~$A_{:,i}$ denotes the i-th column of $A$. For a positive semidefinite matrix $\Lambda$, $\|x - y\|_{\Lambda}^2 = (x - y)^\top \Lambda (x-y)$.  $\R_{>0}$ denotes the set of positive real number whereas $\R_{\geq 0}$ is the set of non-negative real numbers. $\C^i$ denotes the class of $i$-th times differentiable functions. The operator $\nabla_\x$ with $\x\in\R^n$ denotes $[\frac{\partial}{\partial x_1},\ldots,\frac{\partial}{\partial x_n}]^\top$.}
\begin{align}
\begin{split}
    \dx&=[J_s(\x)-R_s(\x)]\nabla_\x H( \x)+G(\x)\u\\
    \y&=G(\x)^\top \nabla_\x H(\x),\label{for:pch}
\end{split}
\end{align}
with the state $\x(t)\in\R^n$ (also called energy variable) at time $t\in\R_{\geq 0}$, the total energy represented by a smooth function $H\colon\R^n\to\R$ called the Hamiltonian, and the I/O ports $\u,\y\in\R^m$. The dynamics depends on a mode $s\in\{1,\ldots,n_s\}$, where $n_s\in\N$ is the total number of system modes.

For each mode $s$, the matrix $J_s\colon\R^n\to\R^{n\times n}$ is skew-symmetric and specifies the interconnection structure and $R_s\colon\R^n\to\R^{n\times n},R=R^\top\succeq 0$ specifies the dissipation in the system. The interchange with the environment is defined by the matrix $G\colon\R^n\to\R^{n\times m}$. The structure of the interconnection matrix~$J_s$ is typically derived from kinematic constraints in mechanical systems, Kirchhoff’s law, power transformers, gyrators, etc. Loosely speaking, the interconnection of the elements in the SPHS is defined by $J_s$, whereas the Hamiltonian $H$ characterizes their dynamical behavior. These are typically imposed by nonlinear elements, physical coupling effects, or highly nonlinear electrical and magnetic fields. The port variables~$\u$ and~$\y$ are conjugate variables in the sense that their duality product defines the power flows exchanged with the environment of the system, for instance, forces and velocities in mechanical systems.
\section{Problem Setting}\label{sec:ps}
We consider the problem of learning the dynamics of a physical switching system with the following assumptions. 
\begin{assum}\label{ass:1}
    We have access to potentially noisy observations $\tilde{\x}(t)\in\R^n$ of the system state $\x(t)\in\R^n$ whose evolution over time $t\in\R_{\geq 0}$ follows the \emph{unknown} dynamics 
\begin{align}
\label{for:pchobs}
        \dot \x&=[J_s(\x)-R_s(\x)]\nabla_\x H( \x)+G(\x)\u,
\end{align} 
where a continuous solution $\x(t)$ exists and is unique for all $x_0\in\X_0\subseteq\R^n$ and $t\in\R_{\geq 0}$. The measurements $\tilde \x(t_i)$ are generated by $\tilde \x(t_i) = \x(t_i) + \bm{\eta}$ where $\bm{\eta}\in\R^n$ is distributed according to a zero-mean Gaussian $\eta\sim\mathcal{N}(\bm{0},\diag[\sigma_1^2,\ldots,\sigma_n^2])$ with unknown variances~ $\sigma_1^2,\ldots,\sigma_n^2\in\R_{\geq 0}$.
\end{assum}
\begin{assum}\label{ass:2}
    The interconnection matrix $J_s\colon\R^n\to\R^{n\times n}$, dissipation matrix $R_s\colon\R^n\to\R^{n\times n}$ and the I/O matrix $G\colon\R^n\to\R^{n\times m}$ are known for all $s\in\{1,\ldots,n_s\}$.
\end{assum}
\begin{assum}\label{ass:3}
    The mode $s$ of~\cref{for:pchobs} is measurable.
\end{assum}
\Cref{ass:1} ensures that we have access to Gaussian noise corrupted state measurements that can be provided by either real sensors or an state observer. Additionally, it is assumed that the real system can be described by an SPHS where the switch configuration $s$ does not entail algebraic constraints on the state variables $\x$, which holds for many switching systems, including contact situations with finite stiffness, see~\citep{van2009state}. \Cref{ass:2} is necessary but not restrictive, as the Hamiltonian $H$ typically captures the unstructured, nonlinear uncertainties of the system, while the structure of $J,R$ and $G$ is rather simple, and accurate estimates can often be easily obtained, see~\citep{van2000l2}. Finally,~\cref{ass:3} allows to measure the mode of the physical switching system, for instance, a hopping robot can be equipped with a contact sensor to distinguish the flying phase and the ground contact phase. 

\begin{prob}
Given~\cref{ass:1,ass:2,ass:3}, a dataset of time\-stamps $\{t_i\}_{i=1}^N$ and noisy state observations with input and mode~$\{\tilde \x(t_i),\bm{u}(t_i),s(t_i)\}_{i=1}^N$ of the system~\cref{for:pchobs}, we aim to learn a probabilistic model
\begin{align}
\label{for:pchmodel}
        \dot \x&=[J_{\hat s}(\x)-R_{\hat{s}}(\x)]\nabla_\x \hat{H}( \x)+G(\x)\u,
\end{align} 
with an estimated Hamiltonian $\hat{H}$ and a switching policy $\hat s\colon \x\mapsto\{1,\ldots,n_s\}$ that allows for uncertainty quantification.
\end{prob}
\section{Gaussian Process Switching Port-Hamiltonian Systems}\label{sec:mod}
\begin{figure}[t]
\begin{center}
\vspace{0.2cm}
	\input{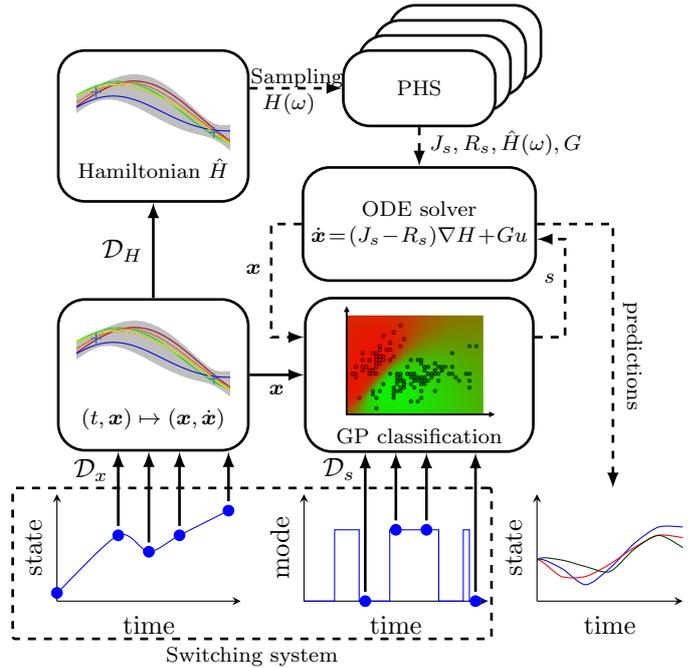}
	\vspace{-0.6cm}\caption{Block diagram of a GP-PHS. The training phase is indicated by solid lines and the prediction phase by dashed lines.}\vspace{0.1cm}
	\label{fig:gpphs}
\end{center}
\end{figure}

In this section, we propose Gaussian process switching Port-Hamiltonian systems (GP-SPHS) whose structure is visualized in~\cref{fig:gpphs}. Starting with data of a physical switching system (dashed rectangle), we use a surrogate GP to estimate the state's derivative, which is fed into another GP to model the Hamiltonian as a nonparametric, probabilistic function. A separate GP classifier is trained on the measured modes of the system. After training, samples of the GP regressor and classifier are drawn. Then, an ODE solver can produce sample trajectories based on the drawn Hamiltonian and the switching policy that allows for uncertainty quantification of the model.

In contrast to parametric models, our approach is beneficial in several ways. First, we do not rely on prior knowledge about the parametric structure of the Hamiltonian and switching policy. For instance, instead of assuming a linear spring model and fitting Hooke's spring constant based on data, the nonparametric structure of the GP allows to learn complex, nonlinear spring models. The switching policy is typically challenging to model as it is affected by the nonlinearities in the system. For instance, switching in an electronic system might be triggered by a threshold voltage of some nonlinear element.

Second, the Bayesian nature of the GP enables the model to represent \emph{all possible} SPHS under the Bayesian prior based on a finite number of data points. This is not only interesting from a model identification perspective, but this uncertainty quantification can also be explicitly useful for robust control approaches.

In the following section, we present the general structure of GP-SPHS starting with the prior model followed by the learning and prediction procedure.
\subsection{Modeling}
First, we start with the modeling of the Hamiltonian $H$ using a GP. Let~$(\Omega, \mathcal{F},P)$ be a probability triple with the probability space~$\Omega$, the corresponding~$\sigma$-algebra~$\mathcal{F}$ and the probability measure~$P$. Then, a GP $\mathcal{GP}(m_{\mathrm{GP}}, k)$ with mean $m_{\mathrm{GP}}(x)$ and kernel $k(\x,\x^\prime)$ is a stochastic process on a set $\mathcal{X} \subseteq \R^n$ where any finite set of points follows a normal distribution. The kernel $k$ is a measure for the correlation of two inputs. Combined with Bayes' Theorem, GPs can provide tractable statistical inference for regression and classification~\citep{rasmussen2006gaussian}.
We place an independent GP prior on each dimension of the derivative of the estimated Hamiltonian
\begin{align}\label{for:GPdh}
    \frac{\partial \hat{H}}{\partial x_i}\sim \GP(0,k_H(\x,\x^\prime)),\quad \forall i\in\{1,\ldots,n\},
\end{align}
where $k_H$ is the squared exponential kernel $k_H(\x,\x^\prime) = \sigma_f^2\exp(-\|\x- \x^\prime\|_{\Lambda}^2)$. This choice of the kernel results in sampled Hamiltonians which are smooth and allows to approximate any continuous function arbitrarily exactly, see~\citep{rasmussen2006gaussian}.
The hyperparameters of the squared exponential kernel are the signal noise $\sigma_f\in\R_{>0}$ and the lengthscales $\Lambda=\diag(l_1^2,\ldots,l_n^2)\in\R_{>0}^{n}$, which can be optimized based on data.
Due to the finite number of system's modes, the switching policy is handled as GP multi-class classification problem. Thus, we put a GP prior on the switching policy $\hat s$ such that
\begin{align}
    \hat{s}\sim \GP(0,k_s(\x,\x^\prime)).
\end{align}
By design, the GP approaches to classification and regression problems are similar, except that the Gaussian distributed error model in the regression case is replaced by a Bernoulli distribution over a logistic function, see~\citep{williams1998bayesian} for more details.

The choice of the kernel function $k_s$ allows to incorporate prior knowledge about the switching policy, e.g., by using linear or polynomial kernels. However, without prior knowledge, the squared exponential kernel is a common choice as a starting point due to its universality property.
\subsection{GP-SPHS Training}
As indicated in the problem formulation, we consider an observed system trajectory
\begin{align} \label{for:dataset0}
\begin{split}
        \mathcal{D}_x&=\{(t_1,\tilde{\x}(t_1), \bm{u}(t_1)),\ldots,(t_{N},\tilde{\x}(t_{N}), \bm{u}(t_N))\}\\
\end{split}
\end{align}
 of the unknown dynamics~\cref{for:pchobs} corresponding to measured inputs $\{\bm{u}(t_1), \ldots, \bm{u}(t_N)\}$, where $N$ is the total number of training pairs consisting of a time $t_i$ and a noisy state measurement $\tilde{\x}(t_i)$. The first challenge we address is how to extract data of the form $\{\x,\dx\}$ from \cref{for:dataset0} so we may apply GP regression to learn the derivative of the Hamiltonian of the system~\cref{for:pchobs}. The advantages over classical filtering techniques are the inherent noise handling and the uncertainty quantification, which we include in the learning of the Hamiltonian. To obtain the derivative $\dx$, we exploit that GPs are closed under affine operations~\citep{adler2010geometry}. We learn $n$ separated GPs on the training sets $\mathcal{D}_x$, one GP for each dimension $j$ of the state $\x\in\R^n$. We rearrange the training set as input and output matrix written as
 \begin{align}
 \begin{split}\label{for:dataset}
     T&=[t_1,\ldots,t_N]\in\R^{1\times N}\\
     \tilde{X}&=[\tilde{\x}(t_1),\ldots,\tilde{\x}(t_N)]^\top\in\R^{N\times n}.
      \end{split}
 \end{align}
 
Using a differentiable kernel $k$, we obtain the distribution for each element of the \emph{state derivative} $\dot{x}_j\in\R$ by
 \begin{align}
     \mu(\dot{x}_j \mid t,\D_x)&=\bm{k}^{(1)}\left(t, T\right)^{\!\top}\!K^{-1} \tilde{X}_{:,j}\label{for:meanvarxdx}\\
     \var(\dot{x}_j \mid t,\D_x)&=\bm{k}^{(1,2)}(t, t)-\bm{k}^{(1)}\left(t, T\right)^{\top}K^{-1} \bm{k}^{(1)}\left(t, T\right),\notag
 \end{align}
 where~\cref{for:meanvarxdx} is the closed-form posterior distribution of the GP, see~\citep{rasmussen2006gaussian} for more details.
The function~$K\colon \R^{N}\times  \R^{N}\to\R^{N\times N}$ is called the Gram matrix whose elements are given by $K_{i,i'}= k(t_{i},t_{i'})+\delta(i,i')\sigma_j^2$ for all $i',i\in\{1,\ldots,N\}$ with the delta function $\delta(i,i')=1$ for $i=i'$ and zero, otherwise. The vector-valued function~$\bm{k}\colon \R\times  \R^{N}\to\R^N$, with the elements~$k_i = k(t,T_{i})$ for all $i\in\{1,\ldots,N\}$, expresses the covariance between an input~$t$ and the input training data $T$. 

The term $k^{(l)}$ denotes the derivative of the kernel function $k$ with respect to the $l$-th argument, i.e., $k^{(1)}(t,T)_i= \frac{\partial}{\partial z}k(z, T_i)\big\vert_{z=t}$, $k^{(2)}(t,T)_i= \frac{\partial}{\partial z}k(t,z)\big\vert_{z=T_i}$, and ${k}^{(1,2)}(t,t)= \frac{\partial^2}{\partial z\partial z^\prime}k(z,z^\prime)\big\vert_{z=t,z^\prime=t}$ for $i=1,\ldots,N$. Further, an estimate of the state $\x(t_i)$ based on the noise state measurement $\tilde{\x}(t_i)$ is obtained by standard GP regression $\mu(x_j \mid t,\D_x)=\bm{k}\left(t, T\right)^{\top}K^{-1} \tilde{X}_{:,j}$. 
With the Gaussian prior $\tilde{X}\vert \bm{\varphi},T\sim\mathcal{N}\left(\bm{0},K \right)$ we can compute the negative log marginal likelihood (NLML) to learn the unknown hyperparameters $\bm{\varphi}=[\sigma_f,l_1,\ldots,l_n,\sigma_1^2,\ldots,\sigma_n^2]^\top\in\R^{2n+1}$ of~\cref{for:meanvarxdx}.
At this point, we have access to the estimated state $\mu(\x \mid t,\D_x)$ and its derivative $\mu(\dx \mid t,\D_x)$.
\begin{rem}
For the sake of simplicity, we focus here on a single trajectory. However, the above procedure can be repeated for multiple trajectories in the same manner.
\end{rem}
Next, the learning of the derivatives of the Hamiltonian~$H$ is presented. Analogously to~\cref{for:meanvarxdx}, each dimension of $\nabla_\x H(\x)$ is learned by an independent GP. For this purpose, we need a suitable set of input and output training data. The input data consists of the state estimates $\mu(\x \mid t,\D_x)$ as $H(\x)$ is state dependent. For the output data, we can rewrite the system dynamics~\cref{for:pchobs} as
\begin{align}
    [J_s(\x)-R_s(\x)]^{-1}(\dot \x-G(\x)\u)&=\nabla_\x H( \x),
\end{align}
to isolate $\nabla_\x H$ on the right-hand site if $[J_s(\x)-R_s(\x)]$ is invertible. However, that might not hold for all states $\x$ and system modes $s$. In this case, the invertable subspace of $[J_s(\x)-R_s(\x)]$ is used such that the number of available training data can differ between the dimensions of $\nabla_\x H$. The output data for the $j$-dimension $\dot{X}^j$ is given by
\begin{align}\label{for:setdX}
     \dot{X}^j=[\bm{p}^j_{s_i}(\bar{\x}_i)(\bar{\dx}_i-G(\bar{\x}_i)\bm{u}(t_i))]_{\forall i\in\mathcal{I}^j},
\end{align}
where $\bar{\x}_i=\mu(\x \mid t_i,\D)$, $\bar{\dx}_i=\mu(\dx \mid t_i,\D)$ and $s_i=s(t_i)$. The vector $\bm{p}^j_{s_i}(\bar{\x}_i)\in\R^{1\times n}$
is defined such that it is a unique solution of $\nabla_{x_j} H(\bar{\x}_i)=[\bm{p}^j_{s_i}(\bar{\x}_i)(\bar{\dx}_i-G(\bar{\x}_i)\bm{u}(t_i))]$ for all $i$ in an index set $\mathcal{I}^j\subseteq\{1,\ldots,N\}$. Loosely speaking, the pair $(\bar{\x}_i,\bar{\dx}_i)$ is exploited as a training point for the $j$-dimension of $\nabla_{x} H$ if there exists a bijective mapping between $(\bar{\x}_i,\bar{\dx}_i)$ and $\nabla_{x_j} H$. The vector $\bm{p}^j_{s_i}(\bar{\x}_i)$ will equal the $j$-th dimension of the inverse of $[J_s(\bar{\x}_i)-R_s(\bar{\x}_i)]$ if it exists. In this case, each index set $\mathcal{I}^j$ contains all training points, i.e., $\mathcal{I}^j=\{1,\ldots,N\}$.
Then, the corresponding input data is defined by
\begin{align}\label{for:setX}
     X^j=[\mu(\x \mid t_i,\D)]_{\forall i\in\mathcal{I}^j}.
\end{align}
Then, we can create a new dataset $\mathcal{D}_H=\{X^j,\dot{X}^j\}_{j=1,\ldots,n}$ that allows to learn an estimated $\nabla_\x\hat H$. 

Analogously to~\cref{for:meanvarxdx}, the Gram matrix $K_H^j\in\R^{N^j\times N^j}$, where $N_j$ is the cardinality of $\mathcal{I}^j$, is defined by
\begin{align}\label{for:kphs}
    K_H^j&=\begin{bmatrix}k_H(X^j_{:,1},X^j_{:,1}) & \ldots & k_H(X^j_{:,1},X^j_{:,N^j})\\\vdots & \ddots & \vdots\\ k_H(X^j_{:,N^j},X^j_{:,1}) & \ldots & k_H(X_{:,N^j},X_{:,N^j})\end{bmatrix}\notag\\
    &+\begin{bmatrix}\var(\dot{x}_j \mid t_1,\D) & 0 & 0\\ 0 & \ddots & 0\\ 0 & 0 & \var(\dot{x}_j \mid t_{N^j},\D)\end{bmatrix}.
\end{align}
Here, we use the posterior variance~\cref{for:meanvarxdx} of the estimated state derivative data~$\dot X$ as noise in the covariance matrix \cref{for:kphs}. This allows us to consider the uncertainty of the estimation in the modeling of the SPHS. Finally, the unknown hyperparameters $\bm{\varphi}_H$ can be computed by minimization of the NLML 
\begin{align}
    -\log \proby(\dot{X}^j\vert \bm{\varphi}_H,X^j)&\sim[\dot{X}^j]^\top [K_H^j]^{-1} \dot{X}^j+\log\vert K_H^j \vert,\label{for:loglik}
\end{align}
for each $j=\{1,\ldots,n\}$ via, e.g., a gradient-based method as the gradient is analytically tractable.

As a last step, the state-dependent switching policy $\hat s$ needs to be learned. For this purpose, we create a training set based on the estimated state and the measured mode of the system, i.e., $\mathcal{D}_s=\{(\bar{\x}_1,s_1),\ldots,(\bar{\x}_N, s_N)\}$. Analogously to the GP model for the estimated Hamiltonian $\hat H$, the hyperparameters of the kernel $k_s$ can be optimized by means of the NLML. However, in case of GP classification, there exists no analytically tractable solution for the NLML but, instead, a numerical approximation can be used, see~\citep[Algorithm 3.1]{rasmussen2006gaussian}.

In~\Cref{alg:cap1}, we summarize the steps to train the GP-SPHS.
\begin{algorithm}
\caption{Training  of GP-SPHS}\label{alg:cap1}
\begin{algorithmic}
\Require Trajectory $\mathcal{D}\leftarrow \{(t_i,\tilde{\x}(t_i))\}_{i=1,\ldots,N}$
\Require Control inputs $\{\bm{u}(t_i)\}_{i=1,\ldots,N}$
\Require System modes $\{s(t_i)\}_{i=1,\ldots,N}$
\State \textbf{Obtain $(\x,\dx)$ pairs}
\State Train $n$ independent GPs with $\mathcal{D}_x$~\cref{for:dataset}
\State Create training set $\mathcal{D}_H=\{X^j,\dot{X}^j\}_{j=1,\ldots,n}$~\cref{for:setX,for:setdX}
\State Create training set $\mathcal{D}_s$
\State \textbf{Obtain GP-SPHS model:}
\State Learn $\nabla_x\hat H$ with $\mathcal{D}_H$ and $\{\bm{u}(t_i)\}_{i=1,\ldots,N}$
\State Compute posterior variance $\var(\dx \mid T,\mathcal{D}_x)$ \cref{for:meanvarxdx}
\State Minimize NLML~\cref{for:loglik} to estimate $\bm{\varphi}_H$
\State Learn $\hat s$ with $\mathcal{D}_s$, minimize NLML
\end{algorithmic}
\end{algorithm}
The complexity of the algorithm is dominated by the cost of training the GPs that is $\mathcal{O}(N^3+N^2n)$ with respect to the number $N$ of training points and the dimension $n$ of the state $\x$. The complexity can be reduced by inducing points methods, for instance, as presented in~\citep{wilson2015kernel}.
\subsection{Prediction}
Once the GP-SPHS is trained, an ODE solver is used to compute trajectories from the estimated dynamics
\begin{align}\label{for:est_dyn}
    \dot \x&=[J_{\hat{s}(\x)}(\x)-R_{\hat{s}(\x)}(\x)]\nabla_\x \hat{H}( \x)+G(\x)\u.
\end{align}
For this purpose, samples from the stochastic Hamiltonian~$\hat H$ and the switching policy $\hat s$ are required. For $\hat H$, we can draw samples\footnote{We use the notation $\hat H(\x^*,\omega)$ to distinguish between a sample and the stochastic process $\hat H(\x^*)$} $\nabla_x \hat H(\x^*,\omega),\forall\omega\in\Omega$ from the posterior distribution using the joint distribution at $\x^*\in\R^n$
\begin{align}
    \begin{bmatrix}\dot{X}^j\\\nabla_{x_j}\hat H(\x^*)\end{bmatrix}\!\!=\!\mathcal{N}\left(\!\bm{0},\!\begin{bmatrix}K_H^j & k_H(X^j,\x^*)\\k_H(X^j,\x^*)^\top & k_H(\x^*,\x^*)\end{bmatrix}\right),\label{for:jointdis}
\end{align}
to obtain an estimate of the Hamiltonian's derivative $\nabla_x H$. For using a numerical integrator to generate trajectories, we will need to be able to access the same sample at an arbitrary number of points at arbitrary locations. However, the joint distribution~\cref{for:jointdis} only allows to sample from a finite number of a-priori known test points $\x^*\in X^*$. Thus, we are looking for an analytically tractable function to approximate a sample $\nabla_x \hat H(\x^*,\omega)$ which can be achieved via the Matheron’s rule, see~\citep{wilson2020efficiently}. The error of the interpolation does not affect the SPHS properties as shown later.

In addition to a sampled Hamiltonian, the switching policy $\hat s$ is required to solve the estimated dynamics~\cref{for:est_dyn}. As predictions with a GP classifier are not analytically tractable, a Laplacian approximation is used to obtain an approximated GP posterior, see~\citep{williams1998bayesian}. Then, analogously to~\cref{for:jointdis}, we can draw from the posterior distribution of the approximated GP to obtain a sample $\hat{s}(\x,\omega),\omega\in\Omega$, see~\citep[Algorithm 3.2]{rasmussen2006gaussian}.

In~\Cref{alg:cap}, we summarize the steps to achieve sampled trajectories of a GP-SPHS starting at $\x_0\in\mathcal{X}_0$.
\begin{algorithm}
\caption{Prediction with GP-SPHS}\label{alg:cap}
\begin{algorithmic}
\Require Number of desired sample trajectories
\Require Initial state $\x_0$
\For{each sample}
\State Sample posterior Hamiltonian $\nabla_x\hat{H}\sim\GP$
\State Sample posterior switching policy $\hat{s}\sim\GP$
\State ODE solver for GP-SPHS~\cref{for:est_dyn} with
\State $\phantom{a}\nabla_x\hat{H}(\x,\omega)$ and $\hat s(\x,\omega)$
\EndFor
\end{algorithmic}
\end{algorithm}
 As data-driven method, the accuracy of GP-SPHS typically increases with the number of training points $N$. The complexity scales with $\mathcal{O}(N^2)$. Thus, the choice of $N$ is a trade-off between computational complexity and accuracy of the prediction. 
Next, we prove that a GP-SPHS generates valid samples of a SPHS with probability 1.
\subsection{Theoretical analysis}\label{sec:Prop}
\begin{thm}\label{prop:1}
    Consider a Hamiltonian $\hat H$ and a switching policy $\hat s$ sampled from a GP-SPHS. For all realizations $\omega$ in the sample space $\Omega$, the dynamics
    \begin{align}
    \begin{split}\label{for:phs}
            \dx&=[J_{\hat{s}(\x,\omega)}(\x)\!-\!R_{\hat{s}(\x,\omega)}(\x)]\nabla_\x \hat{H}( \x,\omega)+G(\x)\u\\
        \y&=G(\x)^\top \nabla_\x \hat H(\x,\omega),
        \end{split}
    \end{align}
    on a compact space $\X\subseteq\R^n$, describes a switching Port-Hamiltonian system that is almost surely passive with respect to the supply rate $\u^\top \y$.
\end{thm}
\begin{pf}
  See \cref{sec:app}.
\end{pf}
As a consequence, the GP-SPHS model allows us to build physically correct models in terms of conversation or dissipation of energy. Next, we show that even for approximations of $\hat H$ and $\hat s$, the resulting SPHS remains passive.
\begin{cor}\label{cor:1}
Consider a GP-SPHS trained on a dataset~\cref{for:dataset0} with sampled Hamiltonian $\hat H(\cdot,\omega)$ and switching policy $\hat s(\cdot,\omega)$ with $\omega\in\Omega$. Let $\hat{H}^*\colon\R^n\to\R$ be a smooth and bounded function approximator of $\hat H(\cdot,\omega)$. Then,~$
\dx=[J_{\hat{s}^*(\x)}(\x)\!-\!R_{\hat{s}^*(\x)}(\x)]\nabla_\x \hat{H}^*( \x)+G(\x)\u$ describes a Port-Hamiltonian system that is passive with respect to the supply rate $\u^\top \y$.
\end{cor}
\begin{pf}
    The proof results from~\cref{prop:1}.
\end{pf}
As a consequence, we can use Matheron’s rule, see~\citep{wilson2020efficiently}, that generates smooth and bounded functions to achieve analytically tractable approximations of $\nabla_\x\hat H(\x^*,\omega)$ and $\hat s(\x^*,\omega)$. To model more complex systems, we often wish to simplify the modeling process by separating the system into connected subsystems. The class of PHS are closed under such interconnections, see~\citep{cervera2007interconnection}. We will show that GP-SPHS share the same characteristic which allows to learn separated subsystems of complex systems.
\begin{prop}\label{prop:inter}
    Consider two GP-SPHS~\cref{for:phs} described by $\{J_{\hat{s},1},R_{\hat{s},1}, \hat H_1,G_1\}$ with input dimension $m_1\in\N$ and $\{J_{\hat{s},2},R_{\hat{s},2}, \hat H_2,G_2\}$ with input dimension $m_2\in\N$, respectively. Let $(\u_1^{\text{c}},\bm{y}_1^{\text{c}})\in\R^{m_{\text{c}}\times m_{\text{c}}}$ and $(\u_2^{\text{c}},\bm{y}_2^{\text{c}})\in\R^{m_{\text{c}}\times m_{\text{c}}}$ be the corresponding input / output pairs of dimension $m_{\text{c}}$ with $m_{\text{c}}\leq \min\{m_1, m_2\}$ for the connection of the two GP-SPHS. Then, the interconnection $\u_1^{\text{c}}= -\bm{y}_2^{\text{c}}$ and $\u_2^{\text{c}}= \bm{y}_1^{\text{c}}$ yields a GP-SPHS.
\end{prop}
\begin{pf}
  See \cref{sec:app}.
\end{pf}
\Cref{prop:inter} shows that the negative feedback interconnection of two GP-SPHS leads again to a GP-SPHS. This is in particular interesting for passivity-based control approaches, e.g.,~\citep{ortega2004interconnection}.
\section{Numerical Evaluation}\label{sec:sim}
\begin{figure}[b]
\begin{center}
	\includegraphics[width=5cm]{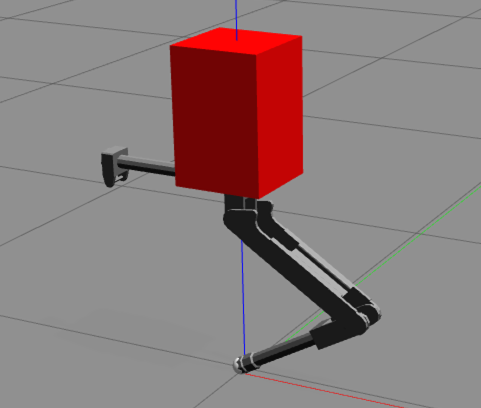}
	\vspace{-0.2cm}\caption{Simulation of a hopper robot in Gazebo (adapted from~\citep{xpp_ros}).}
	\label{fig:hopper}
\end{center}
\end{figure}

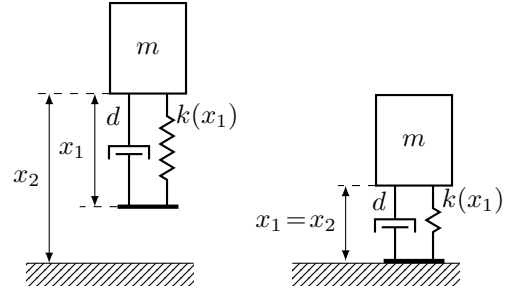
\begin{figure}[tb]
\begin{center}
	\vspace{0.2cm}\begin{tikzpicture}[every node/.style={draw,outer sep=0pt,thick}]
\tikzstyle{spring}=[thick,decorate,decoration={zigzag,pre length=0.3cm,post length=0.3cm,segment length=6}]
\tikzstyle{damper}=[thick,decoration={markings,  
  mark connection node=dmp,
  mark=at position 0.5 with 
  {
    \node (dmp) [thick,inner sep=0pt,transform shape,rotate=-90,minimum width=15pt,minimum height=3pt,draw=none] {};
    \draw [thick] ($(dmp.north east)+(2pt,0)$) -- (dmp.south east) -- (dmp.south west) -- ($(dmp.north west)+(2pt,0)$);
    \draw [thick] ($(dmp.north)+(0,-5pt)$) -- ($(dmp.north)+(0,5pt)$);
  }
}, decorate]
\tikzstyle{ground}=[fill,pattern=north east lines,draw=none,minimum width=0.75cm,minimum height=0.3cm]

\node (M) [minimum width=1cm, minimum height=1.2cm] {$m$};
\node (wall) [ground, minimum width=2.2cm,yshift=-3cm,xshift=-0.5cm] {};
\draw (wall.north east) -- (wall.north west);
\draw [spring] (M.south) ++(0.25cm,0) -- node[draw=none,right,pos=0.2] {$k(x_1)$} +(0,-1.5cm);
\draw [damper] (M.south) ++(-0.25cm,0) -- node[draw=none,left,pos=0.2] {$d$} +(0,-1.5cm);
\draw [ultra thick] (M.south) ++(0.4cm,-1.5cm) -- +(-0.8cm,0);
\draw [dashed] (M.south west) -- +(-1cm,0);
\draw [latex-latex] (M.south west) ++(-0.8cm,0) -- node[left, draw=none,pos=0.5] {$x_2$} +(0,-2.25cm);
\draw [dashed] (M.south) ++(-0.4cm,-1.5cm) -- +(-0.5cm,0);
\draw [latex-latex] (M.south west) ++(-0.2cm,0) -- node[left, draw=none,pos=0.5] {$x_1$} +(0,-1.5cm);

  \begin{scope}[shift={(3.5cm,0)}]
    \node (M) [minimum width=1cm, minimum height=1.2cm,yshift=-1.22cm] {$m$};
\node (wall) [ground, minimum width=2.2cm,yshift=-3cm,xshift=-0.5cm] {};
\draw (wall.north east) -- (wall.north west);
\draw [spring] (M.south) ++(0.25cm,0) -- node[draw=none,right,pos=0.2] {$k(x_1)$} +(0,-1cm);
\draw [damper] (M.south) ++(-0.25cm,0) -- node[draw=none,left,pos=0.2] {$d$} +(0,-1cm);
\draw [ultra thick] (M.south) ++(0.4cm,-1cm) -- +(-0.8cm,0);
\draw [dashed] (M.south west) -- +(-0.5cm,0);
\draw [latex-latex] (M.south west) ++(-0.4cm,0) -- node[left, draw=none,pos=0.5] {$x_1\!=\!x_2$} +(0,-1cm);
  \end{scope}
\end{tikzpicture}
	\caption{Model of hopper robot without ground contact $s=0$ (left) and with ground contact $s=1$ (right).}\vspace{-0.0cm}
	\label{fig:pogo}
\end{center}
\end{figure}

For the evaluation of the proposed GP-SPHS, we aim to learn the dynamics of a hopper robot, see~\cref{fig:hopper}, adapted from~\citep{xpp_ros}. 
For simplification, we assume that the robot is stabilized such that the pose of the body can be fully described by its position on the vertical axis. In this setting, the robot acts as a mechanical system with dissipation. The mass of the robot is known to be $m=\SI{1}{\kilogram}$ and is concentrated mainly in the body of the robot. The stiffness and damping of the two joints are unknown. \Cref{fig:pogo_system} shows one trajectory of a hopper robot that starts above the ground such that the foot is not in contact. During ground contact, the distance between the robot's body to the ground and the robot's body to the foot are identical. For the modeling with a GP-SPHS, we consider the following SPHS from~\citep{van2000l2}
\begin{align}
        \dx&=\begin{bmatrix}\label{for:hopper}
            \frac{\hat{s}-1}{d} & 0 & \hat{s}\\ 0 & 0 & 1\\-\hat s & -1 & -\hat{s}d
        \end{bmatrix}\frac{\partial \hat{H}}{\partial\x}( \x)
\end{align}
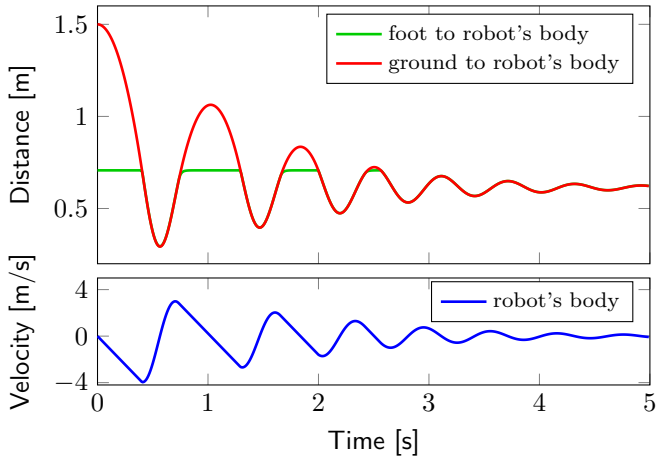
\begin{figure}[b]
\begin{center}
	\tikzsetnextfilename{pogo_system}
\begin{tikzpicture}
\begin{axis}[
  name=plot1,
  xticklabel=\empty,
  ylabel={\sffamily Distance [m]},
  legend pos=north west,
  width=\columnwidth,
  height=5cm,
  ymin=0.2,
  ymax=1.6,
  xmin=0,
  xmax=5,
  legend style={font=\footnotesize},
  legend cell align={left},
  legend pos=north east]
\addplot[color=green!80!black,line width=1pt,no marks] table [x index=0,y index=1]{data/traj_org.csv};
\addplot[color=red,line width=1pt,no marks] table [x index=0,y index=2]{data/traj_org.csv};
\legend{foot to robot's body,ground to robot's body};
\end{axis}
\begin{axis}[
name=plot2,
   at={($(plot1.below south east)+(0,0.5cm)$)}, anchor=above north east,
    xlabel={\sffamily Time [s]},
  ylabel={\sffamily Velocity [m/s]},
  legend pos=north west,
  width=\columnwidth,
  height=3cm,
  ytick={-4,0,4},
  ymin=-4.2,
  ymax=5,
  xmin=0,
  xmax=5,
  legend style={font=\footnotesize},
  legend cell align={left},
  legend pos=north east]
\addplot[color=blue,line width=1pt,no marks] table [x index=0,y index=3]{data/traj_org.csv};
\legend{robot's body};
\end{axis}
\end{tikzpicture} 
	\caption{One trajectory of the hopper robot. During ground contact, the distance between the robot's body to the ground and robot's body to the foot are identical.}\vspace{-0.2cm}
	\label{fig:pogo_system}
\end{center}
\end{figure}
with the length $x_1$ of a (nonlinear) spring, the position of the body's center of mass $x_2$ and its momentum $x_3$ as depicted in~\cref{fig:pogo}. By comparing the trajectory in~\cref{fig:pogo_system} to a linear mass-spring-damper system, we estimate the damping coefficient in~\cref{for:hopper} to $d=2$. For simplicity of demonstration, no external inputs are considered. The contact situation is described by a variable $s=\{0,1\}$ with values $s = 0$ (no contact) and $s = 1$ (contact).
For GP-SPHS training, we performed 20 simulations between $t=[0,5]\si{\second}$ with random initial points $\x_0$ drawn from a uniform distribution over $[0,1]\times[0,2]\times[-1,1]$ such that 
the robot always starts in a non-contact situation. Within each run, every $\SI{0.1}{\second}$ a data point, which includes the time, state and contact situation of the robot, is collected, which leads to a total of 1000 points. The measurements are corrupted by Gaussian noise with a signal-to-noise ratio of $39$dB, $34$dB, and $18$dB for the three states, respectively.

\begin{figure}[b]
\begin{center}
	\includegraphics[width=0.95\columnwidth]{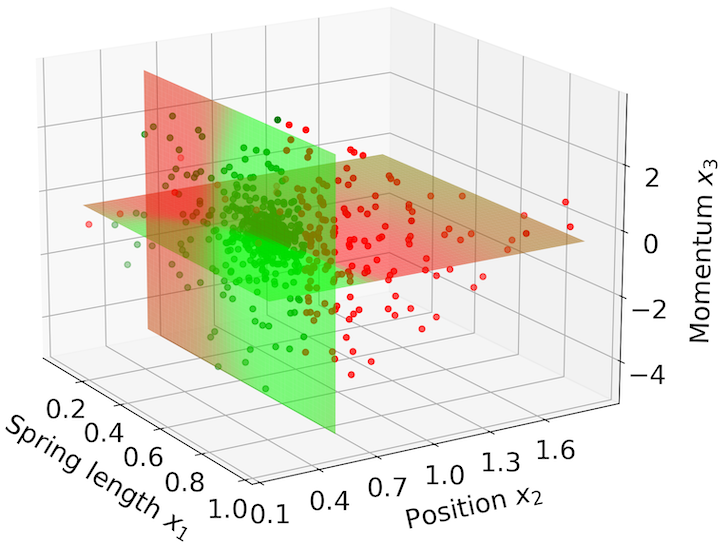}
	\caption{GP classifier for the mode of the system. Based on the training data (points), the classifier estimates for each state the probability of being in mode $s=0$ (no contact, red) and $s=1$ (contact, green).}\vspace{-0.3cm}
	\label{fig:GP_class}
\end{center}
\end{figure}

\begin{figure}[t]
\begin{center}
	\tikzsetnextfilename{pogo_gp1}
\begin{tikzpicture}
\begin{axis}[
  name=plot1,
  ylabel={\sffamily Position $x_2$},
  legend pos=north west,
  width=\columnwidth,
  height=4.3cm,
  xmin=0,
  xmax=3,
  xticklabel=\empty,
  legend style={font=\footnotesize},
  legend cell align={left}]
\addplot[color=blue,line width=1pt,no marks] table [x index=0,y index=2]{data/traj_learned0.csv};
\addplot[color=red,line width=1pt,no marks] table [x index=0,y index=2]{data/traj_learned2.csv};
\addplot[color=green!80!black,line width=1pt,no marks] table [x index=0,y index=2]{data/traj_learned4.csv};
\addplot[color=black,dashed,line width=1pt,no marks] table [x index=0,y index=2]{data/traj_test.csv};
\end{axis}
\begin{axis}[
name=plot2,
   at=(plot1.below south east), anchor=above north east,
  ylabel={\sffamily Contact},
  legend pos=north west,
  width=\columnwidth,
  height=3.6cm,
  ymin=-0.1,
  ymax=1.1,
  xmin=0,
  xmax=3,
  xticklabel=\empty,
  ytick={0,1},
  yticklabels={\sffamily no,\sffamily yes},
  legend style={font=\footnotesize},
  legend cell align={left}]
\addplot[color=blue,line width=1pt,no marks] table [x index=0,y index=4]{data/traj_learned0.csv};
\addplot[color=red,line width=1pt,no marks] table [x index=0,y index=4]{data/traj_learned2.csv};
\addplot[color=green!80!black,line width=1pt,no marks] table [x index=0,y index=4]{data/traj_learned4.csv};
\addplot[color=black,dashed,line width=1pt,no marks] table [x index=0,y index=4]{data/traj_test.csv};
\end{axis}
\begin{axis}[
name=plot3,
   at={($(plot2.below south east)+(0,0.3cm)$)}, anchor=above north east,
  xlabel={\sffamily Time [s]},
  ylabel={\sffamily Hamiltonian $H$},
  legend pos=north west,
  width=\columnwidth,
  ytick={1},
  yticklabels={1},
  height=4.3cm,
  ymax=1.1,
  xmin=0,
  xmax=3,
  legend style={row sep=-8pt},
    legend style={font=\footnotesize},
  legend pos=north east,
  legend style={/tikz/every even column/.append style={column sep=0.2cm}},
  legend columns=3,transpose legend]
\addplot[color=blue,line width=1pt,no marks] table [x index=0,y index=5]{data/traj_learned0.csv};
\addplot[color=red,line width=1pt,no marks] table [x index=0,y index=5]{data/traj_learned2.csv};
\addplot[color=green!80!black,line width=1pt,no marks] table [x index=0,y index=5]{data/traj_learned4.csv};
\addplot[color=white,line width=1pt,no marks] coordinates {(-100,0) (-100,0)};
\addplot[color=black,dashed,line width=1pt,no marks] table [x index=0,y index=5]{data/traj_test.csv};
\addplot[color=white,line width=1pt,no marks] coordinates {(-100,0) (-100,0)};
\legend{\phantom{a},GP-SPHS samples,\phantom{a},\phantom{a}, actual system,\phantom{a}};
\end{axis}
\end{tikzpicture} 
	\vspace{-0.4cm}\caption{Three realizations of the GP-SPHS and the resulting system trajectories. Real system in dashed black. Top: Position of mass always converges to a steady state. Middle: The contact time varies with the realization. Bottom: Hamiltonian is always decreasing over time such that all realizations represent physical plausible systems.}\vspace{0.1cm}
	\label{fig:mss}
\end{center}
\end{figure}
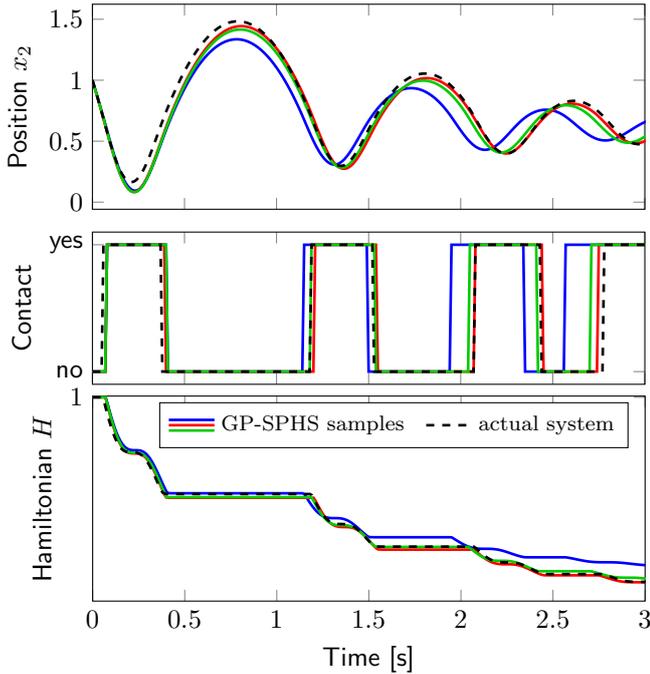

Following~\Cref{alg:cap1}, a GP-SPHS is trained on a MacBook M1 Pro implemented in Python with a runtime of $\SI{22}{\second}$.~\Cref{fig:GP_class} shows the result of the GP classifier for the switching policy $\hat{s}$ based on the training data (circles). For each state $\x$, the classifier is able to predict the probability for each mode of the system. The mean accuracy on the given training data is $0.98$, which indicates that almost all points can be correctly classified. For visualization, two slices in the 3d-space display the predicted probabilities.
The green area indicates the subspace where $s=1$ (contact) and red indicates $s=0$ (no contact). After training, three sample trajectories for an unseen initial point $\x_0$ are drawn based on~\Cref{alg:cap} with Euler's method as an ODE solver and a step size of $\SI{1}{\milli\second}$.~\Cref{fig:mss} depicts the resulting trajectories (solid lines) and the trajectory of the actual system (dashed). The mean squared error between the actual system trajectory and the samples of the GP-PHS is $0.193$. It can be observed that the uncertainty of the prediction is increasing over time, but all realizations represent physical plausible systems for all time as the energy is always decreasing. 
\section{Conclusion}
We propose Gaussian process switching Port-Hamiltonian systems to learn the dynamics of physical switching systems with partially unknown dynamics. The GP-SPHS model guarantees physical plausibility of its sampled trajectories, high flexibility for the learning of nonlinearities and allows for uncertainty quantification. Furthermore, it is shown that GP-SPHS preserve the passivity and interconnection properties of Port-Hamiltonian systems. In future work, we will extend GP-SPHS to a broader class of switching systems and perform experimental validation.


\bibliography{mybib}             

\begin{thebibliography}{27}
\providecommand{\natexlab}[1]{#1}
\providecommand{\url}[1]{\texttt{#1}}
\providecommand{\urlprefix}{URL }
\expandafter\ifx\csname urlstyle\endcsname\relax
  \providecommand{\doi}[1]{doi:\discretionary{}{}{}#1}\else
  \providecommand{\doi}{doi:\discretionary{}{}{}\begingroup
  \urlstyle{rm}\Url}\fi

\bibitem[{Adler(2010)}]{adler2010geometry}
Adler, R.J. (2010).
\newblock \emph{The geometry of random fields}.
\newblock SIAM.

\bibitem[{Anderson et~al.(2020)Anderson, Marshall, L’Afflitto, and
  Dotterweich}]{anderson2020model}
Anderson, R.B., Marshall, J.A., L’Afflitto, A., and Dotterweich, J.M. (2020).
\newblock Model reference adaptive control of switched dynamical systems with
  applications to aerial robotics.
\newblock \emph{Journal of Intelligent \& Robotic Systems}, 100(3), 1265--1281.

\bibitem[{Beckers and Hirche(2016)}]{beckers:cdc2016}
Beckers, T. and Hirche, S. (2016).
\newblock Equilibrium distributions and stability analysis of {G}aussian
  process state space models.
\newblock In \emph{Proceedings of the IEEE Conference on Decision and Control
  (CDC)}, 6355--6361.

\bibitem[{Beckers et~al.(2022)Beckers, Seidman, Perdikaris, and
  Pappas}]{9992733}
Beckers, T., Seidman, J., Perdikaris, P., and Pappas, G.J. (2022).
\newblock Gaussian process port-{H}amiltonian systems: Bayesian learning with
  physics prior.
\newblock In \emph{2022 IEEE 61st Conference on Decision and Control (CDC)},
  1447--1453.

\bibitem[{Bhouri and Perdikaris(2022)}]{bhouri2021gaussian}
Bhouri, M.A. and Perdikaris, P. (2022).
\newblock Gaussian processes meet {NeuralODEs}: a {B}ayesian framework for
  learning the dynamics of partially observed systems from scarce and noisy
  data.
\newblock \emph{Philosophical Transactions of the Royal Society A}, 380(2229),
  20210201.

\bibitem[{Brogliato and Brogliato(1999)}]{brogliato1999nonsmooth}
Brogliato, B. and Brogliato, B. (1999).
\newblock \emph{Nonsmooth mechanics}, volume~3.
\newblock Springer.

\bibitem[{Cervera et~al.(2007)Cervera, van~der Schaft, and
  Ba{\~n}os}]{cervera2007interconnection}
Cervera, J., van~der Schaft, A.J., and Ba{\~n}os, A. (2007).
\newblock Interconnection of port-{H}amiltonian systems and composition of
  dirac structures.
\newblock \emph{Automatica}, 43(2), 212--225.

\bibitem[{Chen et~al.(2018)Chen, Rubanova, Bettencourt, and
  Duvenaud}]{conf/nips/ChenRBD18}
Chen, R.T.Q., Rubanova, Y., Bettencourt, J., and Duvenaud, D.K. (2018).
\newblock Neural ordinary differential equations.
\newblock In \emph{Advances in Neural Information Processing Systems},
  volume~31.

\bibitem[{Desai et~al.(2021)Desai, Mattheakis, Sondak, Protopapas, and
  Roberts}]{desai2021port}
Desai, S.A., Mattheakis, M., Sondak, D., Protopapas, P., and Roberts, S.J.
  (2021).
\newblock Port-{H}amiltonian neural networks for learning explicit
  time-dependent dynamical systems.
\newblock \emph{Physical Review E}, 104(3), 034312.

\bibitem[{Greydanus et~al.(2019)Greydanus, Dzamba, and
  Yosinski}]{greydanus2019hamiltonian}
Greydanus, S., Dzamba, M., and Yosinski, J. (2019).
\newblock {H}amiltonian neural networks.
\newblock \emph{Advances in Neural Information Processing Systems}, 32.

\bibitem[{Hou and Wang(2013)}]{hou2013model}
Hou, Z.S. and Wang, Z. (2013).
\newblock From model-based control to data-driven control: Survey,
  classification and perspective.
\newblock \emph{Information Sciences}, 235, 3--35.

\bibitem[{Jiahao et~al.(2021)Jiahao, Hsieh, and
  Forgoston}]{jiahao2021knowledgebased}
Jiahao, T.Z., Hsieh, M.A., and Forgoston, E. (2021).
\newblock Knowledge-based learning of nonlinear dynamics and chaos.
\newblock \emph{Chaos: An Interdisciplinary Journal of Nonlinear Science},
  31(11), 111101.

\bibitem[{Karniadakis et~al.(2021)Karniadakis, Kevrekidis, Lu, Perdikaris,
  Wang, and Yang}]{karniadakis2021physics}
Karniadakis, G., Kevrekidis, I., Lu, L., Perdikaris, P., Wang, S., and Yang, L.
  (2021).
\newblock Physics-informed machine learning.
\newblock \emph{Nature Reviews Physics}, 3(6), 422--440.

\bibitem[{Maschke and van~der Schaft(1993)}]{maschke1993port}
Maschke, B.M. and van~der Schaft, A.J. (1993).
\newblock Port-controlled {H}amiltonian systems: modelling origins and
  systemtheoretic properties.
\newblock In \emph{Nonlinear Control Systems Design}, 359--365. Elsevier.

\bibitem[{Nageshrao et~al.(2015)Nageshrao, Lopes, Jeltsema, and
  Babu{\v{s}}ka}]{nageshrao2015port}
Nageshrao, S.P., Lopes, G.A., Jeltsema, D., and Babu{\v{s}}ka, R. (2015).
\newblock Port-{H}amiltonian systems in adaptive and learning control: A
  survey.
\newblock \emph{IEEE Transactions on Automatic Control}, 61(5), 1223--1238.

\bibitem[{Ortega and Garcia-Canseco(2004)}]{ortega2004interconnection}
Ortega, R. and Garcia-Canseco, E. (2004).
\newblock Interconnection and damping assignment passivity-based control: A
  survey.
\newblock \emph{European Journal of Control}, 10(5), 432--450.

\bibitem[{Rasmussen and Williams(2006)}]{rasmussen2006gaussian}
Rasmussen, C.E. and Williams, C.K. (2006).
\newblock \emph{{Gaussian} processes for machine learning}.
\newblock MIT press Cambridge.

\bibitem[{Rath et~al.(2021)Rath, Albert, Bischl, and von
  Toussaint}]{rath2021symplectic}
Rath, K., Albert, C.G., Bischl, B., and von Toussaint, U. (2021).
\newblock Symplectic {Gaussian} process regression of maps in {H}amiltonian
  systems.
\newblock \emph{Chaos: An Interdisciplinary Journal of Nonlinear Science},
  31(5), 053121.

\bibitem[{Ridderbusch et~al.(2021)Ridderbusch, Offen, Ober-Bl{\"o}baum, and
  Goulart}]{ridderbusch2021learning}
Ridderbusch, S., Offen, C., Ober-Bl{\"o}baum, S., and Goulart, P. (2021).
\newblock Learning {ODE} models with qualitative structure using {Gaussian}
  processes.
\newblock In \emph{Proceedings of the IEEE Conference on Decision and Control},
  2896--2896.

\bibitem[{Van Der~Schaft and Camlibel(2009)}]{van2009state}
Van Der~Schaft, A. and Camlibel, M.K. (2009).
\newblock A state transfer principle for switching port-{H}amiltonian systems.
\newblock In \emph{Proceedings of the IEEE Conference on Decision and Control
  (CDC)}, 45--50.

\bibitem[{Van~der Schaft(2000)}]{van2000l2}
Van~der Schaft, A. (2000).
\newblock \emph{L2-gain and passivity techniques in nonlinear control}.
\newblock Springer.

\bibitem[{Van Der~Schaft and Jeltsema(2014)}]{van2014port}
Van Der~Schaft, A. and Jeltsema, D. (2014).
\newblock Port-{H}amiltonian systems theory: An introductory overview.
\newblock \emph{Foundations and Trends in Systems and Control}, 1(2-3),
  173--378.

\bibitem[{Williams and Barber(1998)}]{williams1998bayesian}
Williams, C.K. and Barber, D. (1998).
\newblock Bayesian classification with {G}aussian processes.
\newblock \emph{IEEE Transactions on pattern analysis and machine
  intelligence}, 20(12), 1342--1351.

\bibitem[{Wilson and Nickisch(2015)}]{wilson2015kernel}
Wilson, A. and Nickisch, H. (2015).
\newblock Kernel interpolation for scalable structured {G}aussian processes
  ({KISS-GP}).
\newblock In \emph{International conference on machine learning}, 1775--1784.
  PMLR.

\bibitem[{Wilson et~al.(2020)Wilson, Borovitskiy, Terenin, Mostowsky, and
  Deisenroth}]{wilson2020efficiently}
Wilson, J., Borovitskiy, V., Terenin, A., Mostowsky, P., and Deisenroth, M.
  (2020).
\newblock Efficiently sampling functions from {G}aussian process posteriors.
\newblock In \emph{International Conference on Machine Learning}, 10292--10302.
  PMLR.

\bibitem[{Winkler(2017)}]{xpp_ros}
Winkler, A.W. (2017).
\newblock {Xpp - A collection of ROS packages for the visualization of legged
  robots}.
\newblock \urlprefix\url{https://doi.org/10.5281/zenodo.1037901}.

\bibitem[{Wu et~al.(2018)Wu, Zhang, Cheng, and Xin}]{wu2018switched}
Wu, X., Zhang, K., Cheng, M., and Xin, X. (2018).
\newblock A switched dynamical system approach towards the economic dispatch of
  renewable hybrid power systems.
\newblock \emph{International Journal of Electrical Power \& Energy Systems},
  103, 440--457.

\end{thebibliography}
\appendix
\section{}\label{sec:app}
\textbf{Proof of~\cref{prop:1}}
    As we place a GP with squared exponential kernel on $\nabla_x\hat H$, all realizations $\nabla_x\hat H(\x,\omega)$ with $\omega\in\Omega$ are smooth functions in $\x$, see~\citep[Section 4.2]{rasmussen2006gaussian}. Using the fact that every smooth function $\hat H: \R^n \to \R$ defines a Port-Hamiltonian vector field under the affine transformation $\hat H \mapsto (J_s-R_s) \nabla \hat H + G u$, see~\citep{maschke1993port}, all realizations $\nabla_x\hat H(\x,\omega)$ define PHS vector fields.\balance 
    
    To show passivity, we must first show that there exists a $c_0\in\mathbb{R}$ such that for almost all $\omega \in \Omega$, $\hat H(\x,\omega)\geq c_0$, see~\citep{9992733}. This guarantees that infinite energy cannot be extracted from the system. As the mean and covariance of a GP with squared exponential kernel are bounded, see~\citep{beckers:cdc2016}, the standard deviation metric $d(\x,\x^\prime)=\var(\hat H(\x)-\hat H(\x^\prime))^{1/2}$ for all $\x,\x^\prime\in\X$ is totally bounded, and we get $d(\x, \x^\prime)< c_1$ for some $c_1 \in \mathbb{R}_{> 0}$. This, together with the sample path continuity resulting from the squared exponential kernel allows us to conclude that $\hat H$ is almost surely bounded on $\X$ as the GP is closed under linear transformations, i.e,~$\Prob(\sup_{\x\in\R^n}\vert \hat H \vert<\infty)=1$. Thus, we have shown that there exists a $c_0 \in \R$ such that almost surely $\hat H > c_0$.  Further, we can now consider the time-derivative of the sampled Hamiltonian for a $\omega\in\Omega$ during a mode $s\in\{1,\ldots,n_s\}$
\begin{align}
    \dot{\hat H}(\x,\omega)&=\nabla_\x^\top \hat{H}(\x,\omega) [J_s(\x)-R_s(\x)]\nabla_\x \hat{H}(\x,\omega)\notag\\
    &+\nabla_\x^\top \hat{H}(\x,\omega) G(\x)\u\label{for:tdham}\\
    &=-\nabla_\x^\top \hat{H}(\x,\omega) R_s(\x)\nabla_\x \hat{H}(\x,\omega)+\u^\top \y\leq \u^\top \y,\notag
\end{align}
where $\u^\top \y$ is the supply rate. As the dissipation matrix~$R_s$ is positive semi-definite by definition, equation \cref{for:tdham} can be simplified to $\dot{\hat H}(\x,\omega)\leq \u^\top \y$. Thus, the change in the total energy of the system $\hat H$ in the mode $s$ is less than the supply rate with the difference of the dissipation energy. 

Finally, consider the state $\x^-\in\R^n$ of an SPHS at a
switching time where the switch configuration of the system changes to $s^+\in\{1,\ldots,n_s\}$. Using the continuity property in~\cref{ass:1} and the smoothness of $\hat H$, the new state $\x^+\in\R^n$ just after the switching time satisfies $\x^-=\x^+$ such that $\hat{H}(\x^-,\omega)=\hat{H}(\x^+,\omega)$. As a consequence, the passivity condition $\dot{\hat H}(\x,\omega)\leq \u^\top \y$ holds for all time.

\textbf{Proof of~\cref{prop:inter}.}
Analogous to~\citep{9992733}, we start with the definition of two GP-PHS. The first system is given by
    \begin{align}
    \begin{split}
            \dx&=\hat J_{R1}(\x)\nabla_\x \hat H_1( \x,\omega_1)+G_1(\x)\u_1\\
\y_1&=G_1(\x)^\top \nabla_\x \hat H_1(\x,\omega_1),
    \end{split}\notag
\end{align}
with Hamiltonian $\hat H_1 \sim\GP(0,k_H(\x,\x^\prime))$, state $\x\in\R^{n_1}$, sample $\omega_1\in\Omega$, modes $\hat{s}_1(\x,\omega_1)\in\{1,\ldots,n_{s,1}\}$, $\hat J_{R1}(\x)=J_{\hat{s}_1(\x,\omega_1),1}(\x)-R_{\hat{s}_1(\x,\omega_1),1}(\x)$, and input/output $\bm{u}_1,\bm{y}_1\in\R^{m_1}$. We separate the I/O matrix $G_1$ into $G_1^{\text{c}}(\x)\in\R^{n_1 \times m_c}$ and $G_1^{\text{ex}}(\x)\in\R^{n_1 \times m_1-m_c}$ such that $G(\x)\bm{u}_1=\hat G_1^{\text{c}}\u_1^{\text{c}}+G_1^{\text{ex}}\u_1^{\text{ex}}$ with external input $\u_1^{\text{ex}}\in\R^{m_1-m_c}$. The output for connection $\y_1^\text{c}$ is given by $\y_1^\text{c}=G_1^{\text{c}}(\x)^\top \nabla_\x \hat H_1(\x,\omega_1)$.\\
Analogously, the second system is defined by
\begin{align}
\begin{split}
    \dot{\bm{\xi}}&=\hat J_{R2}(\bm{\xi})\nabla_\bm{\xi} \hat H_2(\bm{\xi},\omega_2)+G_2(\bm{\xi})\u_2\\
\y_2&=G_2(\bm{\xi})^\top \nabla_\bm{\xi} \hat H_2(\bm{\xi},\omega_2),
\end{split}\notag
    \end{align}
with Hamiltonian $\hat H_2\sim\GP(0,k)_H(\bm{\xi},\bm{\xi}^\prime))$, state $\bm{\xi}\in\R^{n_2}$, sample $\omega_2\in\Omega$, modes $\hat{s}_2(\x,\omega_2)\in\{1,\ldots,n_{s,2}\}$, $\hat J_{R2}(\x)=J_{\hat{s}_2(\x,\omega_2),2}(\x)-R_{\hat{s}_2(\x,\omega_2),2}(\x)$, and input/output $\bm{u}_2,\bm{y}_2\in\R^{m_2}$. The I/O matrix $G_2$ is separated into $G_2^{\text{c}}(\bm{\xi})\in\R^{n_2 \times m_c}$ and $\hat G_2^{\text{ex}}(x)\in\R^{n_2 \times m_2-m_c}$ such that $G(\bm{\xi})\bm{u}_2=G_2^{\text{c}}\u_2^{\text{c}}+G_2^{\text{ex}}\u_2^{\text{ex}}$ with $\u_2^{\text{ex}}\in\R^{m_2-m_c}$. The output of the connection $\y_2^\text{c}$ is given by $\y_2^\text{c}=G_2^{\text{c}}(\bm{\xi})^\top \nabla_\bm{\xi} \hat H_2(\bm{\xi},\omega_2)$.\\
For the interconnection $\u_1^{\text{c}}= -\bm{y}_2^{\text{c}}$ and $\u_2^{\text{c}}= \bm{y}_1^{\text{c}}$, we get
    \begin{align}
        \underbrace{\begin{bmatrix} \dx\\ \dot{\bm{\xi}}\end{bmatrix}}_{\bm{z}}&\!\!=\!\!\underbrace{\begin{bmatrix} \hat J_{R1}(\x) & \!\!\!\!\!-G_1^{\text{c}}(\x)[G_2^{\text{c}}(\bm{\xi})]^\top \\ G_2^{\text{c}}(\bm{\xi})[G_1^{\text{c}}(\x)]^\top & \hat J_{R2}(\bm{\xi})\end{bmatrix}}_{J_{\hat{s}(\bm{z},\bm\omega)}(\bm{z})-R_{\hat{s}(\bm{z},\bm\omega)}(\bm{z})}\!\!\begin{bmatrix} \nabla_\x \\ \nabla_\bm{\xi}\end{bmatrix} \!\hat{H}(\bm{z},\bm{\omega})\notag\\
        &+\underbrace{[\hat G_1^{\text{ex}}(\x),\hat G_2^{\text{ex}}(\bm{\xi})]}_{G(\bm{z})}\begin{bmatrix}\u_1^{\text{ex}}\\\u_2^{\text{ex}}\end{bmatrix}\notag\\
        \bm{y}&=[\hat G_1^{\text{ex}}(\x),\hat G_2^{\text{ex}}(\bm{\xi})]^\top\begin{bmatrix} \nabla_\x \\ \nabla_\bm{\xi}\end{bmatrix} \!\hat{H}(\bm{z},\bm{\omega})\notag
    \end{align}
with $\bm{z}=[\x^\top,\bm{\xi}^\top]^\top$, $\bm{\omega}=[\omega_1,\omega_2]^\top$, $J_{\hat{s}(\bm{z},\bm{\omega})}(\bm{z}),R_{\hat{s}(\bm{z},\bm{\omega})}(\bm{z})\in\R^{n\times n},n=n_1+n_2$, and $G(\bm{z})\in\R^{n\times m},m=m_1+m_2-2m_c$ and output $\y\in\R^n$. Then, we can write the derivative of the Hamiltonian of the overall system as $\nabla_\bm{z}H(\bm{z},\bm\omega)=[\nabla_\x^\top\hat H_1(\x,\omega_1),\nabla_\bm{\xi}^\top\hat H_2(\bm{\xi},\omega_2)]^\top$ and the switching variable as $\hat{s}(\bm{z})\in\{1,\ldots,n_{s,1}n_{s,2}\}$ that completes the proof.

\end{document}